\documentclass[aps,psfig,epsfig,preprint] {revtex4}
\usepackage{epsfig}
\usepackage{color}
\usepackage{feynmp}
\usepackage{subfigure}

\newcommand {\sla}[1]{ #1 \!\!\!/}
\begin{document}
\draft
\title{
{\normalsize \hskip4.2in USTC-ICTS-06-07} \\{\bf Backward Compton
Scattering in Strong Uniform Magnetic Field}}
\author{Wei Huang$^{a}$}
\author{Wang Xu$^{a,b}$}
\author{ Mu-Lin Yan$^{a}\footnote{Email address: mlyan@ustc.edu.cn}$}

\affiliation{\centerline{$^a$ Interdisciplinary Center for
Theoretical Study, Department of Modern Physics}
\centerline{University of Science and Technology of China,
 Hefei,Anhui 230026, China}
\centerline{$^b$ Shanghai Institute of Applied Physics, Chinese
Academy of Sciences, Shanghai, China }}
\begin{abstract}
In strong uniform magnetic field, the vacuum Non-Commutative Plane
(NCP) caused by the lowest Landau level(LLL) effect and the QED with
NCP (QED-NCP) are studied. Being similar to the theory of Quantum
Hall effect, an effective filling factor $f(B)$ is introduced to
character the possibility that the electrons stays on LLL. The
backward Compton scattering amplitudes of QED-NCP are derived, and
the differential cross sections for the process with polarized
initial electrons and photons are calculated. The existing
Spring-8's data has been analyzed primitively and some hints for
QED-NCP effects are shown. We propose to precisely measure the
differential cross sections of the backward Compton scattering in
perpendicular magnetic field experimentally, which may lead to
reveal the effects of QED-NCP.

\vskip0.2in


\end{abstract}
\maketitle
\section{introduction}
The physics related to the lowest Landau level (LLL) and the
corresponding non-commutative quantum field theory (NCQFT) have long
been studied  with considerable interests. Considering a charged
particle in a uniform magnetic field, the non-relativistical
Lagrangian is
\begin{equation}\label{1}
L={1\over 2}m(\dot{x}^2 +\dot{y}^2 +\dot{z}^2) +{e\over c}(
\dot{x}A_x +\dot{y}A_y +\dot{z}A_z).
\end{equation}
Here $\nabla \times \overrightarrow{A}=B\hat{y}$, and the gauge is
chosen as $\overrightarrow{A}=(0, 0, -xB)$. Then, solving the
corresponding Schr\"{o}dinger equation, we get the energy
eigenvalues of the Landau Levels\cite{LL}:
\begin{equation}\label{2}
E_{n}=\hbar {eB\over mc}(n+ {1\over 2}).
\end{equation}
Since the separation between the energy levels is
$\mathcal{O}(B/m)$, when the magnetic field is very strong, the
separation becomes very large, and consequently only the lowest
Landau level (LLL) is relevant. From Eq.(\ref{1}), the LLL
Lagrangian reads
\begin{equation}\label{3}
L_{LLL}=-{e\over c}Bx\dot{z}-V(x,z),
\end{equation}
and then by quantum principle we have
\begin{equation}\label{4}
p_z\equiv{\partial L_{LLL}\over \partial \dot{z}}=-{eB\over c}x
~\Rightarrow ~\left[-{eB\over c}x, z\right]=-i\hbar ~\Rightarrow
~\left[ x, z\right]=i{\hbar c\over eB}\equiv i\theta_L,
\end{equation}
where $\theta_L= {\hbar c\over eB}$ is a  parameter to describe the
non-commutativity between space coordinates originated from LLL. It
is essential that the last equation in (\ref{4}) indicates that
there is a Non-Commutative Plane $(x,\;z)$ (NCP) in the
3-dimensional space under very strong $B$ field. The NCP is
perpendicular to the external magnetic field $B$. Nevertheless, we
should note that the physical meaning of the NCP here is {\it
irrelevant} with that in string/M theory \cite{1,2,3}, which belongs
to the Planck scale physics, even though the mathematic formulations
for both of them are similar.

The existence of NCP has been widely used to discuss the quantum
Hall effect and relevant topics in the condensed matter physics and
the mathematical physics \cite{Girviv, PH, Read, Susskind, DJ, AAP}.
In such discussions on quantum Hall effects, the non-commutative
parameter for the NCP is usually taken to be
\begin{equation}\label{5}
\theta=f\theta_L,
\end{equation}
where factor $f=f(\nu, B)$ is a function of filling factor $\nu$ and
$B$ field.

A nature question arisen from the condensed matter physics
discussions mentioned above is whether such sort of NCP discussions
can be extended into the vacuum QED. As a matter of fact (see
Ref.\cite{WY}), the anomalous deviation of (g-2)-factor of muon to
the prediction of the standard model has been attributed to the loop
effects of the QED with NCP , i.e., a kind of Non-Commutative QED.
That could be thought as a rough estimation of NCP effects in QED at
loop level. There are some uncertainties in such (g-2) studies both
due to theoretical treatment errors and due to experiment measure
errors. An exploration to NCP effects in QED at tree level in the
accelerator experiments could be essential to make the thing clear.
The motivation of this letter is to pursue the backward Compton
scattering process in the strong magnetic field which is a QED
process on NCP at tree level, and to explore whether NCP effects
exist or not.

At the first glance, since $e\gamma$-Compton scattering at low
energy is a typical quite well understanding process in QED and has
been widely studied for more than 80 years, it seems hopeless to get
any new subtle information from it nowadays. However, to the best of
our knowledge, the $e\gamma$-scattering inside a perpendicular
strong magnetic field haven't been studied precisely in experiments
until now, therefore such a study may reveal some signal of NCP
effects in QED.

The point for revealing NCP effects caused by the LLL effect in a
process is that the external perpendicular magnetic field $B$
``felt'' by the electron with {\it non-relativistical} motion should
be very strong. We would like to address that the backward Compton
scattering experiment can finely satisfy this precondition. The
backward Compton scattering is a process that the soft laser photon
is back scattered by high energy electron elastically. In the
$e\gamma$-mass center frame (CM), the motion of the electron is
non-relativistic, the Lorentz factor to the laboratory frame is very
large and the magnetic field ``felt'' by the electron
$B=B_{CM}=\gamma B_{Lab}$ becomes very large even if $B_{Lab}$ is
small. For instance, in the beamline BL38B2 in Spring-8 accelerator
with 8GeV electron and 0.01eV photon, the velocity of the electron
$v_{CM}\simeq 0.0006\ll 1,\;\gamma\simeq 15645.6,\; B_{Lab}\simeq
0.117T,\; B_{CM}\simeq 1827T$. It finely satisfies the precondition,
hence the NCP due to LLL can be described by constructing a
non-commutative quantum theory in the mass center frame.

In this letter, we are going to derive the differential cross
section of the backward Compton scattering in a uniform magnetic
field. The QED with NCP will be constructed and employed. We expect
that a precise measurement of this differential cross section will
lead to distinguish the prediction of QED with NCP from the
prediction of QED without NCP.

\section{QED with Non-Commutative Plane}
Non-Commutative Quantum Field Theory (NCQFT) was formulated several
years ago by considering a definite limit of string theory with a
nonzero background ``magnetic'' field \cite{1,2,3}. In natural units
$\hbar=c=1$, the Lagrangian of Non-Commutative QED (NCQED) is
\begin{equation}\label{6}
\mathcal{L}=-{1\over 4}F_{\mu\nu}*F^{\mu\nu}+
\overline{\psi}*(i\gamma^\mu D_\mu-m)*\psi,
\end{equation}
with \begin{equation}\label{7}
D_\mu=\partial_\mu-ieA_\mu,~~~F_{\mu\nu}=\partial_\mu A_\nu-
\partial_\nu A_\mu -ie[A_\mu \c{*} A_\nu],
\end{equation}
and $*$ means the Moyal product:
\begin{equation}\label{8}
(f*g)(x)=\lim_{\xi,\eta\rightarrow 0}\left[e^{\partial_\xi^\mu
\theta_{\mu\nu} \partial_\eta^\nu} f(x+\xi)g(x+\eta)\right],
\end{equation}
where
\begin{equation}\label{9}
[\hat{x}_\mu, \hat{x}_\nu]=i\theta_{\mu\nu}=i\theta C_{\mu\nu},~~~
\theta=f\theta_L=f{1\over eB},
\end{equation}
\begin{equation}\label{10}
C_{\mu\nu}=\left(
             \begin{array}{cccc}
             0 & c_{01} & c_{02} & c_{03} \\
             -c_{01} &      0 & c_{12} & -c_{13} \\
             -c_{02} & -c_{12} &     0 & c_{23} \\
             -c_{03} & c_{13} & -c_{23} &  0
             \end{array} \right),
\end{equation}
and $f=f(B)$ could be thought as an effective filling factor to be
determined experimentally.

We are interested in the NCQED whose non-commutative behaving only
emerges on a NCP, i.e., the QED with NCP (QED-NCP). In this letter,
we calculate in the mass center frame, in which the motion of the
electron is non-relativistic. In the laboratory frame the direction
of the external magnetic field is $\hat{y}$. By means of the Lorentz
transformation, in the mass center frame the electron feels an
electric field along $-\hat{x}$ and a magnetic field along
$\hat{y}$. The electric field has no influence to the
non-commutativity caused by the LLL \cite{DJ}, so that $c_{0i}=0$.
The magnetic field is along $\hat{y}$ and the NCP takes
$(x,\;z)$-plane, so that $c_{13}=1$ and other $c_{ij}=0$.

Note that when one investigated the inverse Compton scattering by
external electromagnetic field or the synchrotron radiation, the
$A_\mu$ in the Lagrangian of QED-NCP should be replaced by $ A_\mu
+A_\mu^{external}$. Because we do not study that process in this
letter, but only interest in the Compton scattering process, the
$A_\mu^{external}$ is neglected.

\section{Backward Compton Scattering in the Magnetic Field}

The backward Compton scattering in the magnetic field is a process
that the soft laser photon is back scattered by high energy electron
elastically. Similar to the existed calculations of Compton
scattering in NCQED\cite{Compton}, we use QED-NCP to calculate the
scattering. The Feynman rules, Feynman diagrams, kinematics and the
differential scattering cross section for backward Compton
scattering are as follows :

\begin{enumerate}

\item From the Lagrangian (\ref{6}), the Feynman rules for QED-NCP are
shown in Fig.\ref{fr:mini:subfig}. The propagators for electron and
photon are the same as QED.
\begin{figure}[!htbp]
\subfigure[$i e\gamma^{\mu}\exp({i p_1\theta p_2/2})$] {
\label{fr:mini:subfig:a}
\begin{minipage}[b]{0.3\textwidth}\centering
\includegraphics*[width=1.5in]{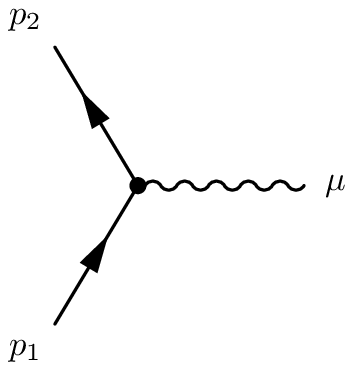}\end{minipage}}%
\hspace{0.5in} \subfigure[$2 e\sin(k_1\theta k_2/2)((k_1-k_2)^\rho
g^{\mu\nu}+(k_2-k_3)^\nu g^{\rho\mu}+(k_3-k_1)^\mu g^{\nu\rho})$] {
\label{fr:mini:subfig:b}
\begin{minipage}[b]{0.4\textwidth}\centering
\includegraphics*[width=1.5in]{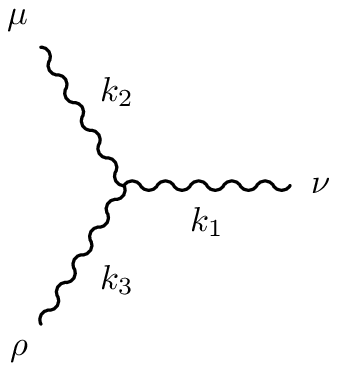}\end{minipage}}
\caption{Feynman rules} \label{fr:mini:subfig}
\end{figure}

\item The Feynman diagrams of $e\gamma$-Compton scattering in QED-NCP
are shown in Fig.\ref{fg:mini:subfig}. $A(i)$ with $i=1,2,3$ denote
the amplitudes of corresponding diagrams. Comparing with that in
QED, there is an additional diagram $A(3)$(see
Fig.\ref{fg:mini:subfig:c}).
\begin{figure}[!htbp]
\subfigure[$A(1)$]{ \label{fg:mini:subfig:a}
\begin{minipage}[b]{0.3\textwidth}\centering
\includegraphics*[width=1.5in]{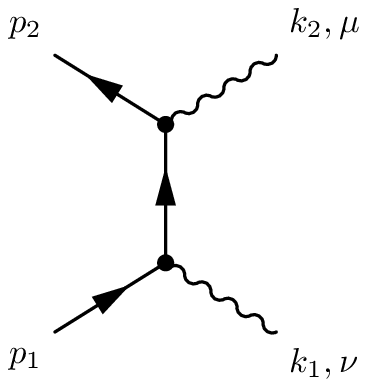}\end{minipage}}%
\subfigure[$A(2)$]{ \label{fg:mini:subfig:b}
\begin{minipage}[b]{0.3\textwidth}\centering
\includegraphics*[width=1.5in]{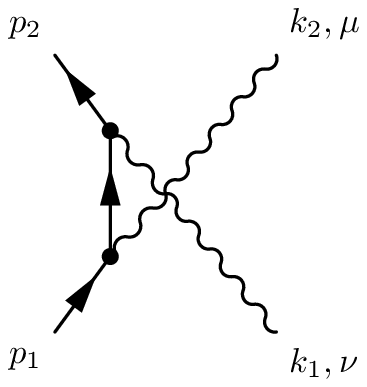}\end{minipage}}%
\subfigure[$A(3)$] { \label{fg:mini:subfig:c}
\begin{minipage}[b]{0.3\textwidth}\centering
\includegraphics*[width=1.5in]{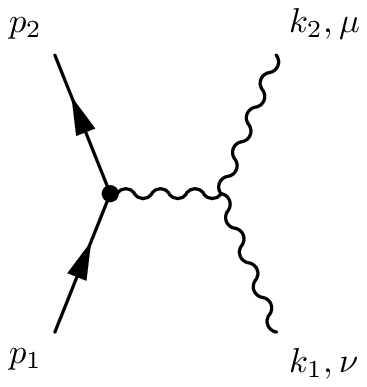}\end{minipage}}
\caption{Feynman diagrams} \label{fg:mini:subfig}
\end{figure}

\item Kinematics (see Fig.\ref{kin:mini:subfig}):

\begin{figure}[!htbp]
\subfigure[The laboratory frame]
{\label{kin:mini:subfig:a}\begin{minipage}[b]{0.5\textwidth}\centering
\includegraphics*[width=2.0in]{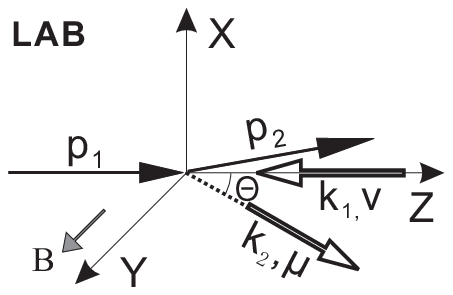}\end{minipage}}%
\subfigure[The mass center frame]
{\label{kin:mini:subfig:b}\begin{minipage}[b]{0.5\textwidth}\centering
\includegraphics*[width=2.0in]{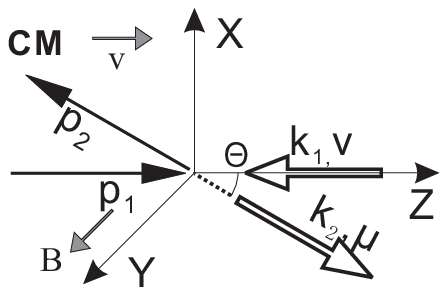}\end{minipage}}
\caption{Kinematics} \label{kin:mini:subfig}
\end{figure}

i) The energies and momenta in the mass center frame :
\begin{eqnarray*}
 && s=(p_1+k_1)^2,~~t=(p_1-p_2)^2,~~u=(p_1-k_2)^2,\\
 && p_1=(\frac{s+m^2}{2\sqrt{s}}, 0, 0, \frac{s-m^2}{2\sqrt{s}} ),\\
 && k_1=\frac{s-m^2}{2\sqrt{s}}(1, 0, 0, -1 ),\\
 && p_2=\frac{s-m^2}{2\sqrt{s}}(\frac{s+m^2}{s-m^2}, -\sin\vartheta\cos\phi,
 -\sin\vartheta\sin\phi, -\cos\vartheta ),\\
 && k_2=\frac{s-m^2}{2\sqrt{s}}(1,\sin\vartheta\cos\phi,
 \sin\vartheta\sin\phi, \cos\vartheta).
\end{eqnarray*}

ii) Polarizations : We are interested in the process with polarized
initial electrons, unpolarized or $\alpha$-polarized initial
photons($\alpha$ is the angle between initial photon polarization
and magnetic field), unpolarized final electrons and unpolarized
final photons. So that the following notations and formulas will be
useful for our goal:
\begin{eqnarray*}
 && {\rm 1)~ initial ~ electron:}\\
 && u^{-1/2}(p_1)\bar{u}^{-1/2}(p_1)\rightarrow\rho=\frac{1}{2}(\sla p_1+m)(1-\gamma^5(-1)
 \gamma^2)\\
 && {\rm 2)~final ~ electron:}\\
 && \sum_i u^i(p_2)\bar{u}^i(p_2)\rightarrow\rho'=\sla p_2+m\\
 && {\rm 3)~photon ~ polarization ~ sums:}\\
 && g_{\mu\nu}=\epsilon^-_\mu\epsilon^{+*}_\nu+\epsilon^-_\mu\epsilon^{+*}_\nu
 -\sum_i\epsilon^T_{i\mu}\epsilon^{T*}_{i\nu}\\
 && \epsilon^+_\mu(k)=( \frac{k^0}{\sqrt{2}|\vec{k}|} ,- \frac{\vec{k}}{\sqrt{2}|\vec{k}|})
 ,~~~ \epsilon^-_\mu(k)=( \frac{k^0}{\sqrt{2}|\vec{k}|} ,
 \frac{\vec{k}}{\sqrt{2}|\vec{k}|})\\
 && {\rm 4)~initial ~ photon:}\\
 && \epsilon^T_{\alpha\mu}=( 0 , \sin\alpha , \cos\alpha , 0), ~~ \epsilon^T_{x\mu}=( 0 , 1 , 0, 0),
 ~~ \epsilon^T_{y\mu}=( 0 , 0 , 1, 0),\\
 && unpolarized : \frac{1}{2}\sum_i\epsilon^T_{i\mu}(k_1)\epsilon^{T*}_{i\mu'}(k_1)\rightarrow\xi_{\mu\mu'}\\
 && \alpha-polarized : \epsilon^T_{\alpha\mu}(k_1)\epsilon^{T*}_{\alpha\mu'}(k_1)\rightarrow\xi_{\mu\mu'}\\
 && {\rm 5)~final ~ photon:}\\
 && unpolarized : \sum_i\epsilon^T_{i\nu}(k_2)\epsilon^{T*}_{i\nu'}(k_2)\rightarrow\xi'_{\nu\nu'}
\end{eqnarray*}

\item Differential cross sections for backward Compton
scattering in QED-NCP are as follows
\begin{equation}\label{11}
\frac{d\sigma}{d\phi d\cos\vartheta}=\frac{e^4}{64\pi^2(s+m^2)}
\xi_{\mu\mu'}\xi'_{\nu\nu'}Tr(\rho'A\rho\bar{A})
\end{equation}
where $A=A(1)+A(2)+A(3)$ and
$A(i)|_{i=1,2,3},~\bar{A}(i)|_{i=1,2,3}$ are :
\begin{eqnarray*}
 &&A(1)= (-1)e^{ip_1\theta p_2/2}e^{ik_1\theta p_2/2}\gamma^\mu\frac{\sla p_1
 +\sla k_1+m}{(p_1+k_1)^2-m^2}\gamma^\nu\\
 &&A(2)= (-1)e^{ip_1\theta p_2/2}e^{-ik_1\theta p_2/2}\gamma^\nu\frac{\sla
 p_1 -\sla k_2+m}{(p_1-k_2)^2-m^2}\gamma^\mu\\
 &&A(3)= (-i)e^{ip_1\theta p_2/2} 2\sin(k_1\theta k_2/2)
 [((k_1+k_2)^\rho g^{\mu\nu}+(k_1-2k_2)^\nu g^{\rho\mu}\\
 &&~~~~~ ~~~~~+(k_2-2k_1)^\mu g^{\nu\rho}]\gamma^\sigma g_{\rho\sigma}/(k_1-k_2)^2\\
 && \bar{A}(1)=(-1)e^{-ip_1\theta p_2/2}e^{-ik_1\theta p_2/2}\gamma^{\nu'}
 \frac{\sla p_1+\sla k_1+m}{(p_1+k_1)^2-m^2}\gamma^{\mu'}\\
 && \bar{A}(2)=(-1)e^{-ip_1\theta p_2/2}e^{+ik_1\theta p_2/2}\gamma^{\mu'}
 \frac{\sla p_1-\sla k_2+m}{(p_1-k_2)^2-m^2}\gamma^{\nu'}\\
 && \bar{A}(3)=(i)e^{-ip_1\theta p_2/2} 2\sin(k_1\theta k_2/2)
 [((k_1+k_2)^{\rho'} g^{\mu'\nu'}+(k_1-2k_2)^{\nu'} g^{\rho'\mu'}\\
 &&~~~~~ ~~~~~+(k_2-2k_1)^{\mu'} g^{\nu'\rho'}]\gamma^{\sigma'} g_{\rho'\sigma'}/(k_1-k_2)^2\\
\end{eqnarray*}

We define the phase factor $\Delta\equiv \frac{k_1\theta
p_2}{2}=-\frac{k_1\theta k_2}{2}$ (notation $k\theta p\equiv
k^\mu\theta_{\mu\nu}p^\nu$), and then the differential cross
sections of the backward Compton scattering with polarized initial
electrons, unpolarized initial photons, unpolarized final electrons
and unpolarized final photons in QED-NCP is :
\begin{eqnarray}\label{12}
\nonumber \frac{d\sigma}{d\phi d\cos\vartheta}& = &
\frac{e^4}{32\pi^2 (s+m^2)} \left((s-m^2)^2+(u-m^2)^2-\frac{4m^2
t(m^4-s u)} {(s-m^2)(u-m^2)}\right)\\  & \times &
\left(-\frac{1}{(s-m^2)(u-m^2)}+\frac{4\sin^2\Delta}{t^2}\right).
\end{eqnarray}
Note that as $m\rightarrow 0$, it coincides with that in NCQED (see
Ref.\cite{Compton}).Note that it's $f(B)$ dependent and goes back to
that in QED when $f(B)\rightarrow 0$. Similarly, for the processes
with any polarization, the differential cross sections could be
calculated, some numerical results are shown in the next section.

\end{enumerate}

\section{Summery and Outlook}

In this letter, the vacuum Non-Commutative Plane (NCP) perpendicular
to the magnetic field and the QED with NCP (QED-NCP) are studied.
Being similar to the theory of Quantum Hall effect, an effective
filling factor $f(B)$ is introduced to character the possibility
that the electrons stays on the lowest Landau level(LLL). The
backward Compton scattering amplitudes of QED-NCP are derived, and
the differential cross sections for the process with polarized
initial electrons and photons are calculated. We propose to
precisely measure the differential cross sections of the backward
Compton scattering in the perpendicular magnetic field
experimentally, which may lead to reveal the effects of QED-NCP.

To show this proposal is practicable, we finally discuss a
measurement of the backward Compton scattering in Spring-8. The
accelerator beamline BL38B2 in Spring-8 has a bending magnet light
source, $10 MeV \gamma$-ray photons are produced in the magnetic
field by backward Compton scattering of FIR laser photons.The energy
of electron in the storage ring is $8GeV$, the perimeter of the ring
is $1436m$, the wavelength of FIR laser photon is $119\mu m$. Then,
in the mass center frame, the Lorentz factor $\gamma\equiv
1/\sqrt{1-v^2}=15645.6$, the magnetic field is $356867eV^2\simeq
1827T$(hence the LLL effect is relevant) and $\theta_L$ is
$9.25\times10^{-6}eV^{-2}\simeq (6{\rm \AA})^2$. We detect final
photon in NCP, i.e., $\phi=\pi$, then the phase factor becomes
$\Delta=f\theta_L\frac{(s-m^2)^2\cos\phi\sin\vartheta}{8s}\simeq
0.4918 f \sin\vartheta$. Substituting all of these into
Eq.(\ref{11}), the realistic calculations are doable. In order to
compare with the data of Spring-8, the results are transferred to
the laboratory frame.

\begin{figure}[!htbp]
\begin{center}
\includegraphics*{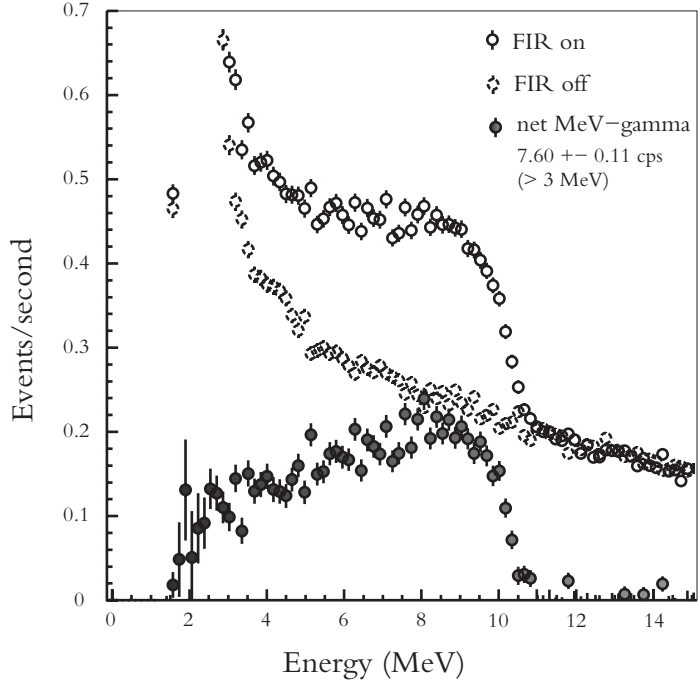}
\caption{Spring-8 data for $e\gamma\rightarrow
e'\gamma'$\cite{mevgamma}. The solid circles indicate the measured
data of ${d\sigma \over dE_\gamma} (E_\gamma)$.}\label{mevgamma}
\end{center}
\end{figure}

Fig.\ref{mevgamma} shows a measurement of the differential cross
section to final photon energy of backward Compton scattering in
Spring-8. The polarization of initial photon is not clear so far,
i.e., it may be unpolarized or linearly polarized. We discuss both
cases as follows:

\begin{enumerate}
\item Unpolarized Initial Photon in Fig.\ref{mevgamma}

Suppose the initial photon is unpolarized, from
Fig.\ref{mevgamma},we can roughly see:
$$\mathcal{R}|_{expt}=
\frac{d\sigma(5MeV)/dE_\gamma}{d\sigma(9MeV)/dE_\gamma}|_{expt}
\simeq\frac{0.15}{0.22} \simeq0.68 .$$

However, the QED prediction is (see Fig.\ref{unpol}):
$$\mathcal{R}|_{QED}=
\frac{d\sigma(5MeV)/dE_\gamma}{d\sigma(9MeV)/dE_\gamma}|_{QED}
\simeq\frac{3.9}{6.2} \simeq0.63 .$$ We find out that $\mathcal{R}
|_{expt}$ is significantly larger than $\mathcal{R} |_{QED}$. A
natural interpretation to this deviation is that the possibility
that the electrons stays on LLL is nonzero, and there is a NCP in
the external magnetic field, which haven't been taken into account
in QED. By means of QED-NCP, and adjusting the effective filling
factor $f(B)$, a suitable $\mathcal{R}|_{QED-NCP}$ can be obtained.
The corresponding prediction with $f(B)=2\times10^{-4}$, which is
consistent with $\mathcal{R}|_{exp}$, are shown in Fig.\ref{unpol} :
$$\mathcal{R}|_{QED-NCP}= \frac{d\sigma(5MeV)/dE_\gamma}{d\sigma(9MeV)
/dE_\gamma}|_{QED-NCP} \simeq\frac{4.3}{6.3}\simeq0.68 .$$

\begin{figure}[!htbp]
\begin{center}
\includegraphics*{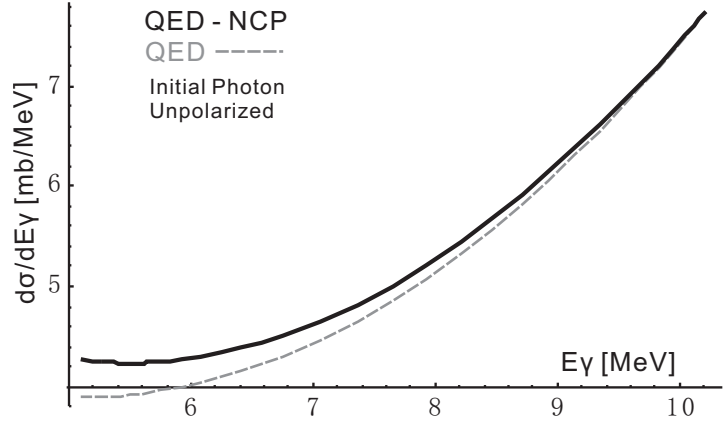}
\caption{Energy dependence of the differential cross section of
unpolarized initial photon.} \label{unpol}
\end{center}
\end{figure}

\begin{figure}[!htbp]
\begin{center}
\includegraphics*{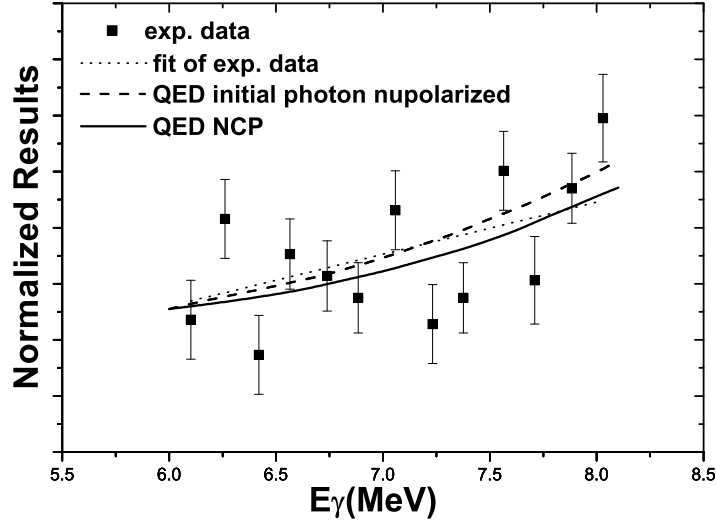}
\caption{Shapes of the experimental data and the QED and QED-NCP
predictions normalized at E$_{\gamma}$ = 6 MeV.} \label{fit}
\end{center}
\end{figure}

However, as we carefully study the shapes of the experimental data
in Fig.\ref{mevgamma} and the QED and QED-NCP predictions in
Fig.\ref{unpol} by normalizing them at E$_{\gamma}$ = 6 MeV,
Fig.\ref{fit} shows that the uncertainties of experimental data are
too large to separate two calculations. In addition, photon
polarization, photon polarization direction, detector
inefficiencies, and radiation corrections due to mirror and windows
will all affect the shapes of experiment data. We realize it is
still too early to decide whether there are QED-NCP effects on this
experimental data of back Compton scattering . A further precise
measurement is needed.

\begin{figure}[!htbp]
\subfigure[$\hat{x}$-polarized initial photon]
{\label{th:mini:subfig:a}\begin{minipage}[b]{0.5\textwidth}\centering
\includegraphics*[width=3.0in]{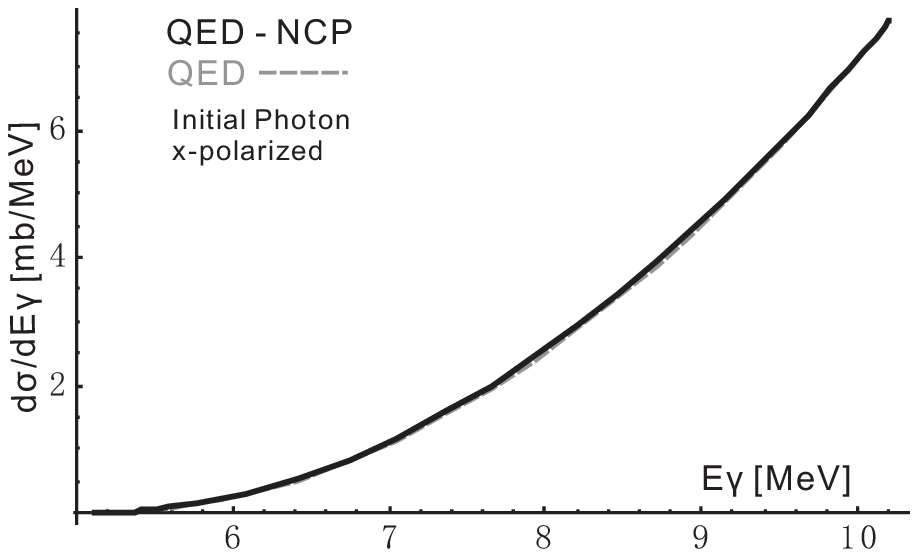}\end{minipage}}%
\subfigure[$\hat{y}$-polarized initial photon]
{\label{th:mini:subfig:b}\begin{minipage}[b]{0.5\textwidth}\centering
\includegraphics*[width=3.0in]{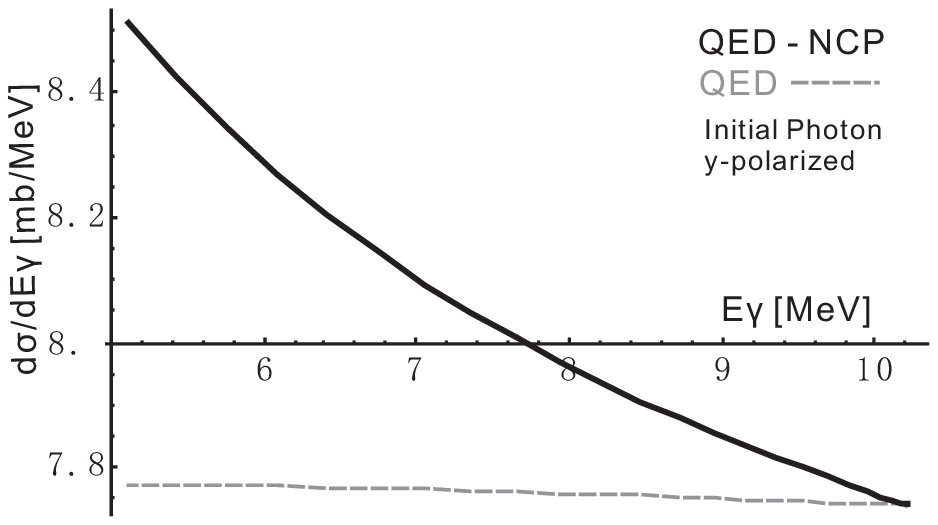}\end{minipage}}
\caption{Energy dependence of the differential cross section of
polarized initial photon.} \label{th:mini:subfig}
\end{figure}

With $f(B)=2\times10^{-4}$, we further consider experiments with
polarized initial photon. The initial laser photons move along
direction $\hat{z}$ and their polarizations are either perpendicular
to or parallel the magnetic field direction $\hat{y}$. The former is
$\hat{x}$-polarized and the energy dependence of differential cross
section in both QED and QED-NCP are shown in
Fig.\ref{th:mini:subfig:a}. We find out that they are very close to
each other. The latter is $\hat{y}$-polarized and the energy
dependence of differential cross section in QED and QED-NCP are very
different (see Fig.\ref{th:mini:subfig:a}). This strongly suggests
that the backward Compton scattering experiment in Spring-8 with
photon polarization parallel the magnetic field is the most
favorable to test the QED-NCP effects.

\item $\alpha$-polarized Initial Photon in Fig.\ref{mevgamma}

Suppose the initial photon in Fig.\ref{mevgamma} is linearly
polarized and the solid angle between the initial photon
polarization and the magnetic field is $\alpha$. With
$\alpha=7\pi/30$, we can roughly fit the QED prediction of the
energy dependence of the differential cross section (see
Fig.\ref{pol:mini:subfig:a}) to the measurement of it in Spring-8
(see Fig.\ref{mevgamma}).Since
$$\mathcal{R}|_{QED}^{\alpha-polarized}=
\frac{d\sigma(5MeV)/dE_\gamma}{d\sigma(9MeV)/dE_\gamma}|_{QED}^{\alpha-polarized}
\simeq\frac{4.3}{6.3} \simeq0.68 ,$$ a more natural interpretation
to the deviation between $\mathcal{R}|_{QED}$ and $\mathcal{R}
|_{expt}$ is that the initial photon is $\alpha$-polarized.

We further consider the ratio of the differential cross section with
$\hat{x}$-polarized initial photon to that with $\hat{y}$-polarized
initial photon : $$\mathcal{R}|_{QED}=
\frac{d\sigma(\hat{x}-polarized)}{d\sigma(\hat{y}-polarized}|_{QED}~,~
\mathcal{R}|_{QED-NCP}=
\frac{d\sigma(\hat{x}-polarized)}{d\sigma(\hat{y}-polarized}|_{QED-NCP}~,$$
and the normalized difference between the QED prediction and the
QED-NCP prediction of the ratio : $$\frac{\mathcal{R}|_{QED}
-\mathcal{R}|_{QED-NCP}}{\mathcal{R}|_{QED}}.$$

The normalized difference shown in Fig.\ref{pol:mini:subfig:b}
suggests that a precise measurement of the backward Compton
scattering in Spring-8 with initial photon polarization
perpendicular to and parallel the magnetic field is still worth to
distinguish the prediction of QED-NCP from that of QED without NCP.

\begin{figure}[!htbp]
\subfigure[QED prediction of the energy dependence of the
differential cross section with $\alpha=\frac{7}{30}\pi$.]
{\label{pol:mini:subfig:a}\begin{minipage}[b]{0.5\textwidth}\centering
\includegraphics*[width=3.0in]{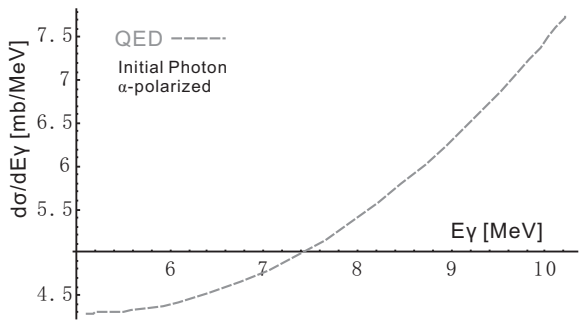}\end{minipage}}%
\subfigure[Energy dependence of the normalized difference of ratio
with $f=2\times10^{-4}$.]
{\label{pol:mini:subfig:b}\begin{minipage}[b]{0.5\textwidth}\centering
\includegraphics*[width=2.5in]{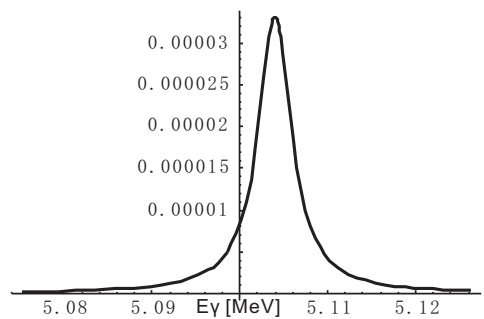}\end{minipage}}
\caption{Suppose the initial photon in Fig.\ref{mevgamma} is
$\alpha$-polarized.} \label{pol:mini:subfig}
\end{figure}

\end{enumerate}

\noindent For both cases, it is practicable to reveal the QED-NCP
effects by means of a precise measurement of backward Compton
scattering in strong uniform magnetic field. It should be
interesting and remarkable.

\begin{acknowledgments}
We would like to acknowledge Prof. M. Fujiwara for discussion. One
of us (MLY) would like to thank Prof. Yong-Shi Wu for helpful
discussions on the quantum Hall effects. This work is supported by
National Natural Science Foundation of China under Grant Numbers
90403021, PhD Program Funds of the Education Ministry of China,
Pujiang Talent Project of the Shanghai Science and Technology
Committee under Grant Numbers 06PJ14114, and Hundred Talent Project
of Shanghai Institute of Applied Physics.
\end{acknowledgments}

\end{document}